\pdfoutput=1
\RequirePackage{ifpdf}
\ifpdf % We are running pdfTeX in pdf mode
\documentclass[pdftex]{sigma}
\else
\documentclass{sigma}
\fi

\numberwithin{equation}{section}

\newcommand\tinyfrac[2]{{\scriptstyle\frac{#1}{#2}}}
\newcommand\n{\vec{n}}
\newcommand\myarg[1]{\left(\n#1\right)}
\newcommand\mye[1]{\vec{e}_{#1}}
\newcommand\myg[1]{\vec{g}_{#1}}
\newcommand\myK[1]{K_{#1}}
\newcommand\myJ[1]{J_{#1}}
\newcommand\myH[1]{H_{#1}}
\newcommand\myI[1]{I_{#1}}
\newcommand\mytH[1]{\dot{H}_{#1}}
\newcommand\mytI[1]{\dot{I}_{#1}}
\newcommand\myf[1]{f_{#1}}
\newcommand\vq{\vec{q}}
\newcommand\vr{\vec{r}}
\newcommand\mysp[2]{\langle #1,\, #2 \rangle}
\newcommand\myttau[2]{\Delta^{M#1}_{N#2}}
\newcommand{\mymatrix}[1]{\mathsf{#1}}
\newcommand\myE{E}
\newcommand\myShift[1]{\mathbb{T}_{#1}}
\newcommand\myShiftInv[1]{\mathbb{T}_{#1}^{-1}}
\newcommand\myShifted[3][0]{%
 \ifcase #1 \mathbb{T}_{#2}#3 \or
 \big( \mathbb{T}_{#2}#3\big) \else ? \fi}

\newtheorem{Theorem}{Theorem}[section]
\newtheorem{Proposition}[Theorem]{Proposition}

\begin{document}

\allowdisplaybreaks

\newcommand{\arXivNumber}{1812.06465}

\renewcommand{\PaperNumber}{028}

\FirstPageHeading

\ShortArticleName{Explicit Solutions for a Nonlinear Vector Model on the Triangular Lattice}
\ArticleName{Explicit Solutions for a Nonlinear Vector Model\\ on the Triangular Lattice}

\Author{V.E.~VEKSLERCHIK}
\AuthorNameForHeading{V.E.~Vekslerchik}

\Address{Usikov Institute for Radiophysics and Electronics, 12 Proskura Str., Kharkiv, 61085, Ukraine}
\Email{\href{mailto:vekslerchik@yahoo.com}{vekslerchik@yahoo.com}}

\ArticleDates{Received December 28, 2018, in final form April 04, 2019; Published online April 13, 2019}

\Abstract{We present a family of explicit solutions for a nonlinear classical vector model with anisotropic Heisenberg-like interaction on the triangular lattice.}


\Classification{37J35; 11C20; 35C08; 15B05}

\section{Introduction}

In this paper, which is a continuation of \cite{V16a,V16b}, we study the possibilities to construct explicit solutions for nonlinear lattices with non-square geometry. Our aim is twofold. First, we want to present an application of some of the ideas and methods elaborated for models on such lattices and even on arbitrary graphs (see, for example, \cite{A98,A01,AS04, BH03,BHS02,BS02,BS10,DNS07}). On the other hand, we want to demonstrate that some of the results obtained for the rectangular lattices can be modified to provide solutions for the models defined on the triangular ones.

We consider a nonlinear vector model defined on the triangular lattice with interaction between the nearest neighbours,
\begin{gather}
 \mathcal{S} = \sum_{ \mathrm{n.n.} } \mathcal{L}(\n_{1},\n_{2})\label{mod:S}
\end{gather}
described by
\begin{gather}
 \mathcal{L}(\n_{1},\n_{2}) = J_{\n_{1},\n_{2}} \ln \big| 1 + K_{\n_{1},\n_{2}} \mysp{ \vq(\n_{1}) - \vq(\n_{2}) }{ \vr(\n_{1}) - \vr(\n_{2}) } \big|\label{mod:L}
\end{gather}
and the restriction
\begin{gather}
 \mysp{ \vq\myarg{} }{ \vr\myarg{} } = 1\label{restr:qr}
\end{gather}
(for all $\n$) where the brackets $\mysp{\;}{}$ stand for the standard scalar product.

From the physical viewpoint, the model \eqref{mod:S} with the Lagrangian \eqref{mod:L}, which can be rewritten, due to the restriction~\eqref{restr:qr}, as
\begin{gather*}
 \mathcal{L}(\n_{1},\n_{2}) = J_{\n_{1},\n_{2}} \ln \big| 1 + K'_{\n_{1},\n_{2}} \mysp{ \vq(\n_{1}) }{ \vr(\n_{2}) } + K'_{\n_{1},\n_{2}} \mysp{ \vq(\n_{2}) }{ \vr(\n_{1}) } \big| + {\rm const},
\end{gather*}
is one of the classical versions of the famous anisotropic Heisenberg spin model of the quantum mechanics~\cite{H82,I82,P87}.

Comparing this work with the previous one, \cite{V16b}, it should be noted that here we study the anisotropic interaction (in~\cite{V16b} the interaction was with $K_{\n_{1},\n_{2}}=1$) on the triangular lattice (instead of the honeycomb one) and take into account the restriction~\eqref{restr:qr}, which is very important for possible physical applications.

In the next section we discuss in more detail the Lagrangian~\eqref{mod:L} and the equations which are the main subject of our study. In Section~\ref{sec:ansatz}, we introduce some auxiliary system and demonstrate how it can used to derive solutions for our equations. These results are used in Section~\ref{sec:solutions} to present three families of the explicit solutions: two types of the $N$-soliton solutions and ones constructed of the determinants of the Toeplitz matrices. Finally, in the conclusion we discuss the limitations of the method we use in this paper and possible continuations of the presented studies.

\section{The model} \label{sec:model}

\begin{figure}[t]\centering
\includegraphics{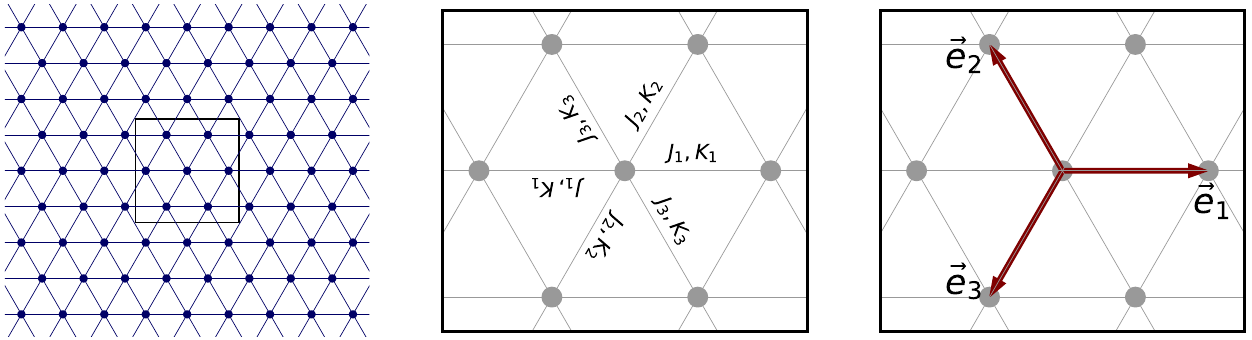}
\caption{The triangular lattice (left), the constants (center) and the base vectors $\mye{i}$ (right).}\label{fig-1}
\end{figure}

It should be noted that, although we study a two-dimensional lattice, we introduce, instead of a pair of basis vectors, \emph{three} coplanar vectors $\vec{e}_{1}$, $\vec{e}_{2}$ and $\vec{e}_{3}$ related by
$ \sum\limits_{i=1}^{3} \vec{e}_{i} = \vec{0} $ and consider the lattice vectors $\n$ as a linear combinations $\vec{n} = \sum\limits_{i=1}^{3} n_{i}\vec{e}_{i}$ (see Fig.~\ref{fig-1}).

In terms of $\{ \mye{i} \}$, the set of the nearest neighbours of a lattice point $\n$ is given by $ \{ \n \pm \mye{i} \}_{i=1,2,3} $ and the discrete action \eqref{mod:S} can be rewritten as
\begin{gather}
 \mathcal{S} = \frac{1}{2} \sum_{ \n } \sum_{i=1}^{3} \sum_{ \varepsilon = \pm 1 } \mathcal{L}( \n, \n + \varepsilon\mye{i} ).\label{mod:Si}
\end{gather}

We cannot solve the most general variant of this model and hereafter impose some restrictions on the coefficients $J_{\n_{1},\n_{2}}$ and $K_{\n_{1},\n_{2}}$ in~\eqref{mod:L}. First we assume that our model is homogeneous: the interaction depends only on the \emph{direction} of $\n_{1} - \n_{2}$,
\begin{gather*}
 J_{\n,\n\pm\mye{i}} = \myJ{i}, \qquad K_{\n,\n\pm\mye{i}} = \myK{i}.
\end{gather*}
Secondly, we assume that constants $\myJ{i}$ are related by
\begin{gather}
 \sum_{i=1}^{3} \myJ{i} = 0\label{restr:J}
\end{gather}
and that
\begin{gather}
 \myK{i} = \frac{ 1 }{ 4 \sinh^{2} \kappa_{i} }, \qquad \sum_{i=1}^{3} \kappa_{i} = 0.\label{restr:K}
\end{gather}
Alternatively, we can formulate these restrictions as
\begin{gather}
 \myJ{i} = \mysp{ \vec{J} }{ \mye{i} }, \qquad \myK{i} = \frac{ 1 }{ 4 \sinh^{2}\mysp{ \vec{K} }{ \mye{i} } }\label{restr:JK}
\end{gather}
with arbitrary vectors $\vec{J}$ and $\vec{K}$. The restriction~\eqref{restr:J} is common for various integrable models. Considering the second restriction, \eqref{restr:K}, we admit that is has no clear physical explanation and is introduced to make equations~\eqref{eq:main-q} and~\eqref{eq:main-r} solvable by the method discussed in what follows rather than to ensure appearance of some additional symmetry of the model or its integrability.

Finally, the big part of the calculations presented below is valid for vectors $\vq,\vr \in \mathbb{R}^{n}$ or $\vq,\vr \in \mathbb{C}^{n}$ with arbitrary $n$. However, as it will be shown below, there arise some problems that, at present, we can solve only in the $n=2$ case. So, we state from the beginning that we consider only two-dimensional real vectors $\vq$ and $\vr$,
\begin{gather*}
 \vq,\vr \in \mathbb{R}^{2}.
\end{gather*}

Looking for the extremals of the action \eqref{mod:Si} under the constraints~\eqref{restr:qr} one arrives at the field equations of our model,
\begin{gather}
 \sum\limits_{i=1}^{3} \myJ{i}\myK{i} \left\{ \frac{ \vq\myarg{+\mye{i}} - \mysp{ \vq\myarg{+\mye{i}} }{ \vr\myarg{} } \vq\myarg{} }{ 1 + \myK{i} \mysp{ \vq\myarg{} - \vq\myarg{+\mye{i}} } { \vr\myarg{} - \vr\myarg{+\mye{i}} } } \right.\nonumber\\
 \left.\qquad {} + \frac{ \vq\myarg{-\mye{i}} - \mysp{ \vq\myarg{-\mye{i}} }{ \vr\myarg{} } \vq\myarg{} }{ 1 + \myK{i} \mysp{ \vq\myarg{} - \vq\myarg{-\mye{i}} } { \vr\myarg{} - \vr\myarg{-\mye{i}} } } \right\} = 0,\label{eq:main-q}\\
 \sum\limits_{i=1}^{3} \myJ{i}\myK{i} \left\{ \frac{ \vr\myarg{+\mye{i}} - \mysp{ \vq\myarg{} }{ \vr\myarg{+\mye{i}} } \vr\myarg{} }{ 1 + \myK{i} \mysp{ \vq\myarg{} - \vq\myarg{+\mye{i}} } { \vr\myarg{} - \vr\myarg{+\mye{i}}} } \right.\nonumber\\
 \left.\qquad {} + \frac{\vr\myarg{-\mye{i}} - \mysp{ \vq\myarg{} }{ \vr\myarg{-\mye{i}} } \vr\myarg{} }{ 1 + \myK{i} \mysp{ \vq\myarg{} - \vq\myarg{-\mye{i}} } { \vr\myarg{} - \vr\myarg{-\mye{i}} } } \right\} = 0.\label{eq:main-r}
\end{gather}

Namely these equations are the main subject of this work. In the following sections we present three families of explicit solutions for the system~\eqref{eq:main-q} and~\eqref{eq:main-r}.

\section{The ansatz}\label{sec:ansatz}

The ansatz that we use in what follows is closely related to the already known ideas that may be termed `star-triangle' (or `star-polygon') transformation or, using the language of, e.g.,~\cite{BS02}, the `three-leg' representation.

We do not start with the form of solutions, which is usual way to introduce an ansatz. Instead, we demonstrate that a wide range of solutions for \eqref{eq:main-q} and \eqref{eq:main-r} can be obtained from a more simple system of equations.

\subsection{Auxiliary system}

The main steps of the proposed ansatz can be described as follows. First, we consider the original lattice $\{ \n \}$ together with its dual, $\{ \n \pm \myg{i} \}$, where new vectors $\myg{i}$ are related to $\mye{i}$ by
\begin{gather}
 \mye{i} = \myg{i+1} - \myg{i-1}.\label{def:g}
\end{gather}
In this equation, as well as in the rest of the paper, we use the following convention: all arithmetic operations with the indices of the vectors $\mye{i}$ and $\myg{i}$ and the constants $\myJ{i}$, $\myK{i}$ and other are understood modulo $3$,
\begin{gather*}
 \mye{i \pm 3} = \mye{i}, \qquad \myg{i \pm 3} = \myg{i}, \qquad \myJ{i \pm 3} = \myJ{i}, \qquad \myK{i \pm 3} = \myK{i}, \qquad \mbox{etc}, \qquad i=1,2,3.%\label{def-indices}
\end{gather*}
In other words, we write $i \pm 1$ bearing in mind the following:
\begin{gather*}
 i+1 :=
 \begin{cases}
2 & i=1, \\
3 & i=2, \\
1 & i=3,
 \end{cases} \qquad
 i-1 :=
 \begin{cases}
3 & i=1, \\
1 & i=2, \\
2 & i=3.
 \end{cases}
\end{gather*}

Then, we introduce a rather simple system of \emph{four-point} equations relating the values of $\vq$ and $\vr$ at the points $\n$, $\n + \myg{i}$ and $\n + \myg{j} + \myg{k}$ and demonstrate that each solution of the latter is at the same time a solution of the field equations~\eqref{eq:main-q} and~\eqref{eq:main-r}. This system can be written as
\begin{gather}
 \myH{i-1} \vq\myarg{+\myg{i-1}} - \myH{i+1} \vq\myarg{+\myg{i+1}} = \frac{ 1 }{ \myJ{i} } F_{i}\myarg{} \vq\myarg{+\myg{i-1}+\myg{i+1}},\nonumber\\
 \myH{i-1} \vr\myarg{+\myg{i+1}} - \myH{i+1} \vr\myarg{+\myg{i-1}} = \frac{ 1 }{ \myJ{i} } F_{i}\myarg{} \vr\myarg{}\label{eq:aux-syst}
\end{gather}
with constant $\myH{i}$ ($i=1,2,3$) and
\begin{gather*}
 F_{i}\myarg{} = \frac{ 1 }{ \myK{i} } + \mysp{ \vq\myarg{+\myg{i+1}} - \vq\myarg{+\myg{i-1}} } { \vr\myarg{+\myg{i+1}} - \vr\myarg{+\myg{i-1}} }
\end{gather*}
(note that structure of $F_{i}\myarg{}$ is that of the denominators of the summands in \eqref{eq:main-q} and \eqref{eq:main-r}). It turns out that the consistency of the system \eqref{eq:aux-syst} implies the following restrictions on the constants $\myJ{i}$:
\begin{gather}
 \myJ{i} = \myI{i+1} - \myI{i-1},\label{eq:J}
\end{gather}
where $\myI{i}$ are arbitrary constants and the following relations between $\myK{i}$ and $\myH{i}$:
\begin{gather}
 \myK{i} = \frac{ \myH{i-1}\myH{i+1} }{(\myH{i-1} - \myH{i+1})^{2} }\label{eq:K}
\end{gather}
(see Appendix \ref{app:a}). Finally, we arrive at the main result of this paper which can be formulated as follows.

\begin{figure}[t]\centering
\includegraphics{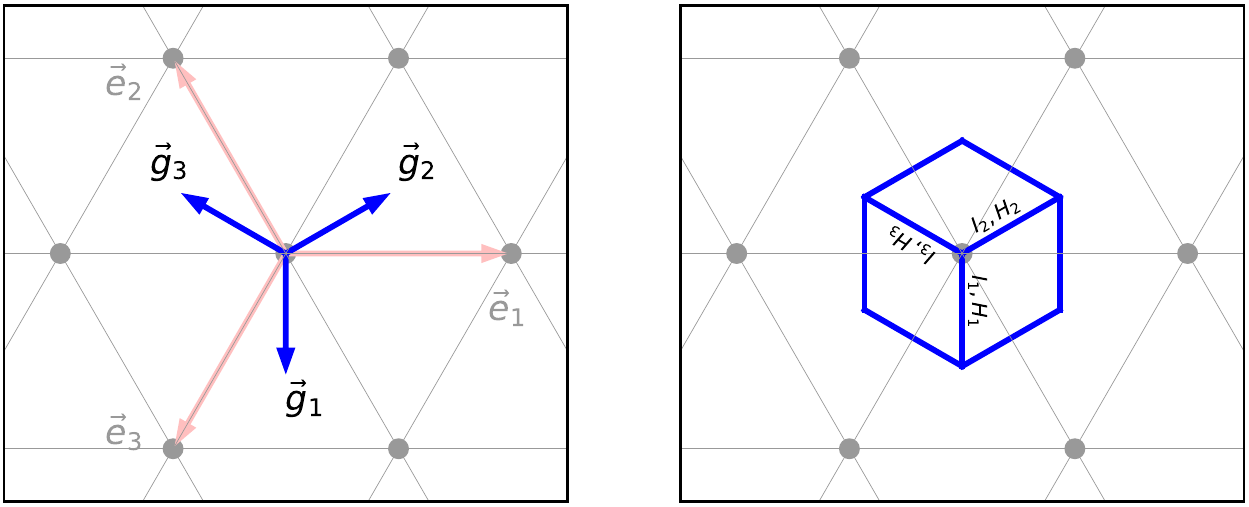}
\caption{ The base vectors $\myg{i}$ (left), and the new constants (right). Each rhombus in the right figure represents the set of nodes involved in the system \eqref{eq:ansatz}.}\label{fig-2}
\end{figure}

\begin{Proposition} \label{prop:one}Each solution of the system
\begin{gather}
 \myH{i-1} \vq\myarg{+\myg{i-1}} - \myH{i+1} \vq\myarg{+\myg{i+1}} = \frac{ 1 }{ \myI{i+1} - \myI{i-1} } F_{i}\myarg{} \vq\myarg{+\myg{i-1}+\myg{i+1}},\nonumber\\
 \myH{i-1} \vr\myarg{+\myg{i+1}} - \myH{i+1} \vr\myarg{+\myg{i-1}} = \frac{ 1 }{ \myI{i+1} - \myI{i-1} } F_{i}\myarg{} \vr\myarg{}\label{eq:ansatz}
\end{gather}
with
\begin{gather}
 F_{i}\myarg{} = \frac{(\myH{i-1} - \myH{i+1})^{2} }{ \myH{i-1}\myH{i+1} } + \mysp{ \vq\myarg{+\myg{i+1}} - \vq\myarg{+\myg{i-1}} } { \vr\myarg{+\myg{i+1}} - \vr\myarg{+\myg{i-1}} }\label{def:F}
\end{gather}
solves equations \eqref{eq:main-q} and \eqref{eq:main-r} with the constants $\myJ{i}$, $\myK{i}$ and the vectors $\mye{i}$ given by \eqref{eq:J}, \eqref{eq:K} and \eqref{def:g}.
\end{Proposition}

We would like to repeat that this proposition describes an ansatz (or reduction): each solution for \eqref{eq:ansatz} with \eqref{def:F} satisfies equations \eqref{eq:main-q} and \eqref{eq:main-r} but the reverse statement is surely not true, only a part of solutions for~\eqref{eq:main-q} and~\eqref{eq:main-r} can be obtained from the system~\eqref{eq:ansatz} with~\eqref{def:F}.

The proof of Proposition~\ref{prop:one} is rather simple except of one non-trivial fact following from the consistency of the system~\eqref{eq:ansatz}: the scalar products $\mysp{ \vq\myarg{+\myg{i}} }{ \vr\myarg{} }$
do not depend on $\n$ (we prove this statement in Appendix~\ref{app:a}).

After this fact is established, the rest of calculations is easy. To compute the summand in~\eqref{eq:main-q},
\begin{gather*}
 \vec{Q}_{\pm i}\myarg{} = \frac{ \myJ{i} }{ \myf{\pm i}\myarg{} } \left\{ \vq\myarg{\pm\mye{i}} - \mysp{ \vq\myarg{\pm\mye{i}} }{ \vr\myarg{} } \vq\myarg{} \right\},
\end{gather*}
where
\begin{gather*}
 \myf{\pm i}\myarg{} = \frac{ 1 }{ \myK{i} } + \mysp{ \vq\myarg{} - \vq\myarg{\pm\mye{i}} } { \vr\myarg{} - \vr\myarg{\pm\mye{i}} }
\end{gather*}
we express $\vq\myarg{\pm\mye{i}} = \vq\myarg{\pm\myg{i+1}\mp\myg{i-1}}$ and $\mysp{ \vq\myarg{\pm\mye{i}} }{ \vr\myarg{} }$ from the first equation from \eqref{eq:ansatz} (translated by $-\myg{i \mp 1}$), note that
$\myf{\pm i}\myarg{}=F\myarg{-\myg{i \mp 1}}$, impose the restriction
\begin{gather}
 \mysp{ \vq\myarg{+\myg{i}} }{ \vr\myarg{} } = \myH{i}\myI{i} \label{eq:U}
\end{gather}
and arrive at
\begin{gather*}
 \vec{Q}_{i}\myarg{} = - \vec{Q}'_{i+1}\myarg{}, \qquad \vec{Q}_{-i}\myarg{} = \vec{Q}'_{i-1}\myarg{}, \qquad i=1,2,3,
\end{gather*}
where
\begin{gather*}
 \vec{Q}'_{i}\myarg{} = \frac{ 1 }{ \myH{i} } \vq\myarg{+\myg{i}} - \myI{i} \vq\myarg{}.
\end{gather*}
Now, one can easily obtain
\begin{gather*}
 \sum_{i=1}^{3}\vec{Q}_{i}\myarg{} = - \sum_{i=1}^{3} \vec{Q}'_{i}\myarg{}, \qquad \sum_{i=1}^{3}\vec{Q}_{-i}\myarg{} = \sum_{i=1}^{3} \vec{Q}'_{i}\myarg{},
\end{gather*}
which means that the right-hand side of \eqref{eq:main-q} vanishes automatically,
\begin{gather*}
 \mbox{r.h.s.~\eqref{eq:main-q}} = \sum_{i=1}^{3} \big[ \vec{Q}_{i}\myarg{} + \vec{Q}_{-i}\myarg{} \big] = 0.
\end{gather*}
Equation \eqref{eq:main-r} can be treated in the similar way.

\subsection{Auxiliary vectors and constants}

One of the ingredients of the proposed method to derive solutions for the field equations \eqref{eq:main-q} and \eqref{eq:main-r} is to use, instead the vectors $\mye{i}$, the vectors $\myg{i}$. It is natural to think of the latter as pointing to the centers of the (three of six) triangular plaquettes adjacent to the point $\n=\vec{0}$. However, it is not necessary to endow the vectors $\myg{i}$ with this geometrical meaning. One can consider them just as three vectors satisfying \eqref{def:g} for given set of~$\mye{i}$. It is easy to show that equation \eqref{def:g} admits one-parametric family of solutions,
\begin{gather}
 \myg{i} = \myg{*} + \tinyfrac{1}{3} ( \mye{i-1} - \mye{i+1} ) \label{rslt:g}
\end{gather}
with arbitrary $\myg{*}$. The fact that the system~\eqref{def:g} does not determine $\myg{i}$ uniquely is not important for our purposes: in the worst case we may have different solutions for different choices of~$\myg{*}$. However, as we see in what follows, this arbitrariness does not affect the final formulae for solutions.

In a similar way, relations \eqref{eq:J} and \eqref{eq:K} considered as \emph{equations} from which one has to find~$\myI{i}$ and~$\myH{i}$ lead to
\begin{gather}
 \myI{i} = \myI{*} + \tinyfrac{1}{3} ( \myJ{i-1} - \myJ{i+1} )\label{rslt:I}
\end{gather}
and
\begin{gather}
 \myH{i} = \myH{*} \exp \big[ \tinyfrac{2}{3}(\kappa_{i-1} - \kappa_{i+1}) \big],\label{rslt:H}
\end{gather}
where $\myI{*}$ and $\myH{*}$ are arbitrary constants or, if one uses~\eqref{restr:JK},
\begin{gather*}
 \myI{i} = I_{*}' + \mysp{ \vec{J} }{ \myg{i} }, \qquad \myH{i} = H_{*}' \exp\big[ 2\mysp{ \vec{K} }{ \myg{i} } \big]
\end{gather*}
with arbitrary $I_{*}'$ and $H_{*}'$.

It should be noted that when we use the three vectors $\mye{i}$ to describe the triangular lattice, instead of a two-vector basis, we bear in mind the restriction $\sum\limits_{i=1}^{3}\mye{i}=0$. This restriction leads to some technical problems when constructing the explicit solutions. Now, if we do not introduce the vectors $\myg{i}$ as pointing to the centers of the plaquettes, but use~\eqref{rslt:g} as the definition, we do not have
the `geometric' restriction: $\sum\limits_{i=1}^{3}\myg{i} \ne 0$. Thus, one of advantages of the proposed construction is that it solves the question of the $\sum\limits_{i=1}^{3}\mye{i}=0$ constraint.

\section{Explicit solutions}\label{sec:solutions}

Here, we derive explicit solutions for the ansatz equations~\eqref{eq:ansatz} and~\eqref{def:F}. To simplify the following formulae and to make the correspondence with the previous works we change the notation. First, we introduce the shifts $\myShift{}$ to indicate the translations:
\begin{gather*}
 \myShift{g_{i}}\vq\myarg{} = \vq\myarg{+\myg{i}}.
\end{gather*}
Then, we rewrite the system of the three equations \eqref{eq:ansatz} as difference/functional equations relating $\vq$ and $\vr$ with $\myShift{\xi}\vq$, $\myShift{\xi}\vr$, $\myShift{\eta}\vq$ and $\myShift{\eta}\vr$ where
$\xi$ and $\eta$ belong to the set of parameters $\{ g_{1},g_{2},g_{3} \}$ that correspond to the translations by the vectors $\{ \myg{1},\myg{2},\myg{3} \}$. This can be done by redefinition of the constants: $\myI{i} \to \mytI{g_{i}}$, $\myH{i} \to \mytH{g_{i}}$. In new terms, equations~\eqref{eq:ansatz} and~\eqref{def:F} become
\begin{gather}
 \mytH{\xi} \myShifted\xi\vq - \mytH{\eta} \myShifted\eta\vq = \frac{ 1 }{ \mytI{\eta} - \mytI{\xi} } F_{\xi,\eta} \myShifted{\xi\eta}\vq, \nonumber\\
 \mytH{\xi} \myShifted\eta\vr - \mytH{\eta} \myShifted\xi\vr = \frac{ 1 }{ \mytI{\eta} - \mytI{\xi} } F_{\xi,\eta} \vr\label{eq:qr}
\end{gather}
and
\begin{gather}
 F_{\xi,\eta} = \frac{ \mytH{\xi} }{ \mytH{\eta} } + \frac{ \mytH{\eta} }{ \mytH{\xi} } - \mysp{ \myShifted\xi\vq }{ \myShifted\eta\vr } - \mysp{ \myShifted\eta\vq }{ \myShifted\xi\vr },\label{eq:F}
\end{gather}
and can be easily bilinearized by the substitution
\begin{gather*}
 \vq = \frac{\vec{\sigma}}{\tau}, \qquad \vr = \frac{\vec{\rho}}{\tau}
\end{gather*}
and
\begin{gather*}
 F_{\xi,\eta} = B_{\xi,\eta} \frac{ \tau \myShifted{\xi\eta}{\tau} } { \myShifted[1]{\xi}{\tau} \myShifted[1]{\eta}{\tau} }
\end{gather*}
with arbitrary constants $B_{\xi,\eta}$, symmetric in $\xi$ and $\eta$ ($B_{\xi,\eta}=B_{\eta,\xi}$), which leads to the bilinear system
\begin{gather}
 \mytH{\xi} \myShifted[1]{\eta}{\tau} \myShifted{\xi}{\vec{\sigma}} - \mytH{\eta} \myShifted[1]{\xi}{\tau} \myShifted{\eta}{\vec{\sigma}} = A_{\xi,\eta} \tau \myShifted{\xi\eta}{\vec{\sigma}},\nonumber\\
 \mytH{\xi} \myShifted[1]{\xi}{\tau} \myShifted{\eta}{\vec{\rho}} - \mytH{\eta} \myShifted[1]{\eta}{\tau} \myShifted{\xi}{\vec{\rho}} = A_{\xi,\eta} \myShifted[1]{\xi\eta}{\tau} \vec{\rho}\label{eq:bilin-sr}
\end{gather}
and
\begin{gather}
 B_{\xi,\eta} \tau \myShifted{\xi\eta}{\tau} + C_{\xi,\eta} \myShifted[1]{\xi}{\tau} \myShifted[1]{\eta}{\tau} = \mysp{ \myShifted{\xi}{\vec{\sigma}} }{ \myShifted{\eta}{\vec{\rho}} }
 + \mysp{ \myShifted{\eta}{\vec{\sigma}} }{ \myShifted{\xi}{\vec{\rho}} },\label{eq:bilin-t}
\end{gather}
where
\begin{gather}
 A_{\xi,\eta} = \frac{ B_{\xi,\eta} }{ \mytI{\eta} - \mytI{\xi} }, \qquad C_{\xi,\eta} = \frac{ \mytH{\xi} }{ \mytH{\eta} } + \frac{ \mytH{\eta} }{ \mytH{\xi} }.\label{eq:bilin-ABC}
\end{gather}
It should be noted that equation \eqref{eq:bilin-t} turns out to be the direct consequence of the equa\-tions~\eqref{eq:bilin-sr}, the definitions \eqref{eq:J} and \eqref{eq:K} together with~\eqref{restr:qr} and \eqref{eq:U} which can be rewritten as
\begin{gather*}
 \mysp{ \vec{\sigma} }{ \vec{\rho} } = \tau^{2}, \qquad \mysp{ \myShifted{\xi}{\vec{\sigma}} }{ \vec{\rho} } = \mytH{\xi} \mytI{\xi} \tau \myShifted[1]{\xi}{\tau}.
\end{gather*}
These equations are, in some sense, an alternative to \eqref{eq:bilin-t} and in the next sections, we will derive solutions for \eqref{eq:qr}--\eqref{eq:F}, and hence for \eqref{eq:ansatz}--\eqref{def:F}, using the following
\begin{Proposition}Any solution for the system
\begin{gather}
 \mytH{\xi} \myShifted[1]{\eta}{\tau} \myShifted{\xi}{\vec{\sigma}} - \mytH{\eta} \myShifted[1]{\xi}{\tau} \myShifted{\eta}{\vec{\sigma}} =
 A_{\xi,\eta} \tau \myShifted{\xi\eta}{\vec{\sigma}},\label{bi:s}\\
 \mytH{\xi} \myShifted[1]{\xi}{\tau} \myShifted{\eta}{\vec{\rho}} - \mytH{\eta} \myShifted[1]{\eta}{\tau} \myShifted{\xi}{\vec{\rho}} = A_{\xi,\eta} \myShifted[1]{\xi\eta}{\tau} \vec{\rho},\label{bi:r}\\
 \mysp{ \vec{\sigma} }{ \vec{\rho} } = \tau^{2},\label{bi:u}\\
 \mysp{ \myShifted{\xi}{\vec{\sigma}} }{ \vec{\rho} } = \mytH{\xi} \mytI{\xi} \tau \myShifted[1]{\xi}{\tau},\label{bi:HI}
\end{gather}
with arbitrary skew-symmetric constants $A_{\xi,\eta}$ $(A_{\xi,\eta} = - A_{\eta,\xi})$ provides solution for the system~\eqref{eq:qr}--\eqref{eq:F} with constants $B_{\xi,\eta}$ and $C_{\xi,\eta}$ given by~\eqref{eq:bilin-ABC} by means of
\begin{gather*}
 \vq = \frac{\vec{\sigma}}{\tau}, \qquad \vr = \frac{\vec{\rho}}{\tau}.
\end{gather*}
\end{Proposition}

Equations \eqref{bi:s} and \eqref{bi:r} are known for many years. In this form, or similar ones, they have appeared in the studies of various integrable systems. For example, the compatibility of, say, equations~\eqref{bi:s} implies that $\tau$ should solve the famous Hirota bilinear difference equation (HBDE)~\cite{H81}, also known as the discrete KP equation,
\begin{gather}
 A_{\xi,\eta}\mytH{\zeta} \myShifted[1]{\xi\eta}{\tau} \myShifted[1]{\zeta}{\tau} - A_{\xi,\zeta}\mytH{\eta} \myShifted[1]{\xi\zeta}{\tau} \myShifted[1]{\eta}{\tau} +
 A_{\eta,\zeta}\mytH{\xi} \myShifted[1]{\eta\zeta}{\tau} \myShifted[1]{\xi}{\tau} = 0.\label{eq:hbde}
\end{gather}
Thus, equations \eqref{bi:s} (or \eqref{bi:r}) can be viewed as the Lax representation of the HBDE (see, e.g., \cite{DJM82a,DJM82b,N97,TSW97,WTS97}). Moreover, it turns out that all components of $\vec{\sigma}$ and $\vec{\rho}$ also solve \eqref{eq:hbde}, thus equations~\eqref{bi:s} and~\eqref{bi:r} can be interpreted as describing the B\"acklund transformations for the HBDE. Also, system~\eqref{bi:s} and~\eqref{bi:r} is closely related to another integrable model: it describes the so-called functional representation of the Ablowitz--Ladik hierarchy~\cite{AL75} (compare~\eqref{bi:s} and~\eqref{bi:r} with equation~(6.10) of~\cite{V02}).

Considering the remaining equations of the auxiliary system, \eqref{bi:u} and \eqref{bi:HI} (or equation~\eqref{eq:bilin-t}), they can be viewed, in the framework of the theory of the HBDE, as a nonlinear restriction, which is compatible with~\eqref{bi:s} and~\eqref{bi:r}. And namely this is the point that makes presented work different from, say,~\cite{V16b}: there we used another way to `close' equations~\eqref{bi:s} and~\eqref{bi:r}, i.e., to relate $\vec{\sigma}$ and $\vec{\rho}$ with the tau-function $\tau$.

Returning to our current task, we would like to repeat that equations \eqref{bi:s}--\eqref{bi:HI} are not new and one does not need to derive some special methods to find their solutions. In what follows we present three families of explicit solutions for \eqref{bi:s}--\eqref{bi:HI}, and, hence, for~\eqref{eq:qr} and~\eqref{eq:F}. In all the cases our approach is to establish the relations between equations we want to solve and the already known equations like the Jacobi identities for the Toeplitz matrices or various identities for the Cauchy-type matrices which gives us a possibility to obtain solutions by means of rather simple calculations.

One of the differences between these cases is different dependence of the constants $\mytI{\xi}$ and~$\mytH{\xi}$ on~$\xi$. To return from the solutions for \eqref{bi:s}--\eqref{bi:HI} to solutions for the field equations~\eqref{eq:main-q} and~\eqref{eq:main-r} we have to invert this dependence: we have to express $\xi$ in terms of $\mytI{\xi}$ and $\mytH{\xi}$, i.e., to obtain the dependence of $g_{i}$ on $\mytI{g_{i}}$ and~$\mytH{g_{i}}$. Then, recalling that $\mytH{g_{i}}=\myH{i}$ and $\mytI{g_{i}}=\myI{i}$ are functions of $\myJ{i}$ and $\myK{i}$ given by~\eqref{rslt:I}, \eqref{rslt:H} and~\eqref{restr:K}, we establish how the parameters $g_{i}$ depend on $\myJ{i}$ and $\myK{i}$.

\subsection{Toeplitz solutions}

The solutions we discuss in this section are built of the determinants of the Toeplitz matrices
\begin{gather}
 \myttau{}{} = \det\left|
 \begin{matrix}
 \alpha_{M} & \alpha_{M+1} & \dots & \alpha_{M+N-1} \\
 \alpha_{M-1} & \alpha_{M} & \dots & \alpha_{M+N-2} \\
 \vdots & \vdots & \ddots & \vdots \\
 \alpha_{M-N+1} & \alpha_{M-N+2} & \dots & \alpha_{M}
 \end{matrix} \right|\label{tp:tau}
\end{gather}
and the shifts defined by
\begin{gather*}
 \myShiftInv{\xi} \alpha_{m} = \alpha_{m} - \xi \alpha_{m+1}.%\label{tp:sh}
\end{gather*}
The action of the shifts can be `translated' to the level of determinants $\myttau{}{}$ as
\begin{gather*}
 \myShiftInv{\xi} \myttau{}{} =
 (-)^{N}
 \det\left| \begin{matrix}
 \xi^{N} & \xi^{N-1} & \dots & 1 \\
 \alpha_{M} & \alpha_{M+1} & \dots & \alpha_{M+N} \\
 \vdots & \vdots & \ddots & \vdots \\
 \alpha_{M-N+1} & \alpha_{M-N+2} & \dots & \alpha_{M+1}
 \end{matrix}
 \right|,\\
 \myShiftInv{\xi\eta} \myttau{}{}
 =
 \frac{1}{\xi-\eta}
 \det\left|
 \begin{matrix}
 \xi^{N+1} & \xi^{N} & \dots & 1 \\
 \eta^{N+1} & \eta^{N} & \dots & 1 \\
 \alpha_{M} & \alpha_{M+1} & \dots & \alpha_{M+N+1} \\
 \vdots & \vdots & \ddots & \vdots \\
 \alpha_{M-N+1} & \alpha_{M-N+2} & \dots & \alpha_{M+2}
 \end{matrix}
 \right|,
\end{gather*}
etc.

Application of the Jacobi identities to the determinants in the above formulae leads to various bilinear identities: the `basic' one,
\begin{gather}\label{tp:A}
 \big( \myttau{}{} \big)^{2} = \myttau{-1}{} \myttau{+1}{} + \myttau{}{-1} \myttau{}{+1}.
\end{gather}
the one-shift,
\begin{gather}\label{tp:Bn}
 0 = \myttau{}{+1} \myShifted{\xi}{ \myttau{}{} } - \myttau{}{} \myShifted{\xi}{ \myttau{}{+1} } + \xi \myttau{-1}{} \myShifted{\xi}{ \myttau{+1}{+1} }
\end{gather}
or
\begin{gather}\label{tp:Bm}
 0 = \myttau{+1}{} \myShifted{\xi}{ \myttau{}{} } - \myttau{}{} \myShifted{\xi}{ \myttau{+1}{} } - \xi \myttau{}{-1} \myShifted{\xi}{ \myttau{+1}{+1} },
\end{gather}
the two-shift,
\begin{gather}\label{tp:Cn}
 0 = (\xi - \eta) \myttau{}{} \myShifted{\xi\eta}{ \myttau{}{+1} } - \xi \myShifted[1]{\xi}{ \myttau{}{+1} } \myShifted[1]{\eta}{ \myttau{}{} } + \eta \myShifted[1]{\xi}{ \myttau{}{} } \myShifted[1]{\eta}{ \myttau{}{+1} },
\\
\label{tp:Cm}
 0 = (\xi - \eta) \myttau{}{} \myShifted{\xi\eta}{ \myttau{+1}{} } - \xi \myShifted[1]{\xi}{ \myttau{+1}{} } \myShifted[1]{\eta}{ \myttau{}{} } + \eta \myShifted[1]{\xi}{ \myttau{}{} } \myShifted[1]{\eta}{ \myttau{+1}{} },
\end{gather}
etc. These identities are the simplest ones, from which one can deduce an infinite set of bilinear identities for the Toeplitz determinants.

Using equations \eqref{tp:A}--\eqref{tp:Cm} and their direct consequences it is easy to demonstrate that functions
\begin{gather*}
 \tau = \myttau{}{},\qquad \vec{\sigma} = \myE \left( \begin{matrix} \myttau{}{+1} \\[1mm] \myttau{+1}{} \end{matrix} \right),\qquad
 \vec{\rho} = \myE^{-1} \left( \begin{matrix} \myttau{}{-1} \\[1mm] \myttau{-1}{}
 \end{matrix} \right),
\end{gather*}
where $\myE$ is the discrete exponential function defined by
\begin{gather*}
 \myShifted{\xi}{\myE} = \mytH{\xi}\mytI{\xi} \; \myE
\end{gather*}
satisfy equations \eqref{bi:s} and \eqref{bi:r} with
\begin{gather*}
 A_{\xi,\eta} = \frac{ \xi - \eta }{ \mytH{\xi}\mytH{\eta} \mytI{\xi}\mytI{\eta} }
\end{gather*}
as well as equations \eqref{bi:u} and \eqref{bi:HI} provided
\begin{gather*}
 \zeta = \mytH{\zeta}^{2} \mytI{\zeta},
\end{gather*}
which means that functions $\vq=\vec{\sigma}/\tau$ and $\vr=\vec{\rho}/\tau$ are solutions for~\eqref{eq:qr} and~\eqref{eq:F}. Thus, to finish the derivation of the solutions for the field equation the only thing we have to do is to rewrite the above formulae in terms of $\n$ bearing in mind that for any function~$f$, $ f\myarg{} = \prod\limits_{i=1}^{3} \big( \myShift{g_{i+1}} \myShiftInv{g_{i-1}} \big)^{n_{i}} f(\vec{0})$ and that in our case $g_{i} = \myH{i}^{2}\myI{i}$ with constants $\myI{i}$ and $\myH{i}$ being defined in~\eqref{rslt:I} and~\eqref{rslt:H}.

The $\n$-dependence of the elements of the determinant \eqref{tp:tau} can be then presented by means of an analogue of the discrete Fourier transform as
\begin{gather*}
 \alpha_{m}\myarg{} = \sum_{\ell=1}^{L} h_{\ell}^{m} \hat{\alpha}_{\ell}\myarg{}
\end{gather*}
with{\samepage
\begin{gather*}
 \hat{\alpha}_{\ell}\myarg{} = \hat{\alpha}_{\ell} \prod_{i=1}^{3} ( 1 - g_{i} h_{\ell} )^{n_{i+1} - n_{i-1}},
\end{gather*}
where $\hat{\alpha}_{\ell}$ and $h_{\ell}$ ($\ell=1,\dots,L$) are arbitrary constants.}

To make the following formulae more clear, we rewrite functions similar to $\myE\myarg{}$ or $\hat{\alpha}(h,\n)$ in the exponential form using the following observation: if some function $f$ satisfies
\begin{gather*}
 \myShifted{g_{i}}f = C_{i} f
\end{gather*}
with positive constants $C_{i}$ then its $\n$-dependence can be written as
\begin{gather*}
 f\myarg{} = \prod\nolimits_{i=1}^{3} \big( \myShift{g_{i+1}} \myShiftInv{g_{i-1}} \big)^{n_{i}} f(\vec{0}) = \prod\nolimits_{i=1}^{3} \left( C_{i+1} / C_{i-1} \right)^{n_{i}} f(\vec{0})
 = f(\vec{0}) e^{ \mysp{ \vec{\phi} }{ \n } }%\label{eq:tn}
\end{gather*}
with
\begin{gather*}%\label{eq:tn-v}
 \vec{\phi} = \frac{2}{3} \sum_{i=1}^{3} \left( \ln \frac{ C_{i+1} }{ C_{i-1} } \right) \mye{i}.
\end{gather*}

Gathering the above formulae and making some trivial calculations we can formulate the main result of this section as follows.

\begin{Proposition}The $N^{\mathrm{th}}$ order Toeplitz solutions for the field equations can be presented as
\begin{gather*}
 \vec{q}\myarg{} = C \frac{ e^{ \mysp{\vec{\varphi}}{\n} } } { \Delta^{0}_{N}\myarg{} } \left( \begin{matrix}
 \Delta^{0}_{N+1}\myarg{} \\[1mm]
 \Delta^{1}_{N}\myarg{}
 \end{matrix} \right), \qquad \vec{r}\myarg{}
 = C^{-1} \frac{ e^{ - \mysp{\vec{\varphi}}{\n} } } { \Delta^{0}_{N}\myarg{} } \left( \begin{matrix}
 \Delta^{0}_{N-1}\myarg{} \\[1mm]
 \Delta^{-1}_{N}\myarg{}
 \end{matrix} \right)
\end{gather*}
where $\myttau{}{}\myarg{}$ $(M=0,\pm 1)$ is the determinant of the Toeplitz matrix
\begin{gather*}
 \myttau{}{}\myarg{} = \det\big| \alpha_{M-j+k}\myarg{}\big|_{j,k=1}^{N},
\end{gather*}
the vector $\vec{\varphi}$ is given by
\begin{gather*}
 \vec{\varphi} = \frac{2}{3} \sum_{i=1}^{3} \left( \ln \frac{ \myH{i+1}\myI{i+1} }{ \myH{i-1}\myI{i-1} } \right) \mye{i}
\end{gather*}
and the function $\alpha_{m}\myarg{}$ is defined as
\begin{gather*}
 \alpha_{m}\myarg{} = \sum_{\ell=1}^{L} \hat{\alpha}_{\ell} h_{\ell}^{m} e^{ \mysp{ \vec{\phi}_{\ell}}{\n} }
\end{gather*}
with
\begin{gather*}
 \vec{\phi}_{\ell} = \frac{2}{3} \sum_{i=1}^{3} \left( \ln \frac{ 1 - \myH{i-1}^{2}\myI{i-1} h_{\ell} } { 1 - \myH{i+1}^{2}\myI{i+1} h_{\ell} } \right) \mye{i}.
\end{gather*}
Here, $\hat{\alpha}_{\ell}$ and $h_{\ell}$ $(\ell=1,\dots ,L)$, $C$ as well as $\myH{*}$ and $\myI{*}$ in the definitions~\eqref{rslt:H} and \eqref{rslt:I} of $\myH{i}$ and $\myI{i}$ $(i=1,2,3)$ are arbitrary constants.
\end{Proposition}

\subsection{`Dark' solitons}

Here we use some of the results of \cite{V14} where we have studied the determinants of the so-called `dark-soliton' matrices
\begin{gather*}
 \tau = \det | \mymatrix{1} + \mymatrix{A} |
\end{gather*}
defined by
\begin{gather*}
 \mymatrix{L} \mymatrix{A} - \mymatrix{A} \mymatrix{L}^{-1} = | 1 \rangle \langle a |,%\label{dark:LA}
\end{gather*}
where $ \mymatrix{L} = \operatorname{diag}( L_{1}, \dots , L_{N})$, $| 1 \rangle$ is the $N$-column with all components equal to~$1$, $\langle a |$ is a $N$-component row that depends on the coordinates describing the model,
and their transformation properties with respect to the shifts defined as $ \myShifted{\xi}{\Omega} = \det | \mymatrix{1} + \myShifted{\xi}{\mymatrix{A}} |$ with
\begin{gather*}
 \myShifted{\xi}{\mymatrix{A}} = \mymatrix{A}\mymatrix{T}_{\xi}%\label{dark:shift-A}
\end{gather*}
and the matrices $\mymatrix{T}$ given by
\begin{gather*}
 \mymatrix{T}_{\xi} = ( \xi - \mymatrix{L}) \big( \xi - \mymatrix{L}^{-1} \big)^{-1}.
\end{gather*}
For our current purposes we need the two facts. First, the determinants $\tau$ satisfy the Fay-like identity
\begin{gather}
( \xi-\eta) \myShifted[1]{\xi\eta}{\tau} \myShifted[1]{\zeta}{\tau} + ( \eta-\zeta ) \myShifted[1]{\eta\zeta}{\tau} \myShifted[1]{\xi}{\tau} + ( \zeta-\xi) \myShifted[1]{\zeta\xi}{\tau} \myShifted[1]{\eta}{\tau} = 0
\label{dark:fay}
\end{gather}
(see \cite{V14}) and, secondly, the superposition of $\myShift{\xi}$ and $\myShift{1/\xi}$ is again the shift, corresponding to zero value of the parameter,
\begin{gather}
 \myShift{\xi} \myShift{1/\xi} = \myShift{0},\label{dark:duality}
\end{gather}
which follows from the corresponding property of the matrices $\mymatrix{T}_{\xi}$, $ \mymatrix{T}_{\xi} \mymatrix{T}_{1/\xi} = \mymatrix{T}_{0}$, that can be verified straightforwardly.

Using only \eqref{dark:fay} and \eqref{dark:duality}, without additional referring to the structure of the matrices~$\mymatrix{A}$, one can demonstrate that vector-functions
\begin{gather*}
 \vec{\sigma} = \left( \begin{matrix}
q_{1} \myE_{1} \myShiftInv{\nu_{1}} \tau\\[1mm]
 q_{2} \myE_{2} \myShiftInv{\nu_{2}} \tau
\end{matrix}
\right),
\qquad
\vec{\rho} = \left(
 \begin{matrix}
 r_{1} \myE_{1}^{-1}
 \myShifted{\nu_{1}}{\tau}\\[1mm]
 r_{2} \myE_{2}^{-1}
 \myShifted{\nu_{2}}{\tau}
\end{matrix}
\right),
\end{gather*}
where $\nu_{1,2}$, $q_{1,2}$ and $r_{1,2}$ are constants related by
\begin{gather}
 \nu_{1}\nu_{2} = 1, \qquad q_{a} r_{a} = \frac{ \nu_{a}^{2} }{ \nu_{a}^{2} - 1 }, \qquad a=1,2, \label{dark:cqr}
\end{gather}
and $\myE_{1,2}$ are two discrete exponential functions, defined by
\begin{gather*}
 \myShifted{\xi}{\myE_{a}} = \frac{ 1 }{ \mytH{\xi} (\xi - \nu_{a}) } \myE_{a}, \qquad a=1,2,%\label{dark:shift-E}
\end{gather*}
satisfy equations \eqref{bi:s} and \eqref{bi:r} with
\begin{gather*}
 A_{\xi,\eta} = (\eta - \xi) \mytH{\xi}\mytH{\eta}
\end{gather*}
as well as equations \eqref{bi:u} and \eqref{bi:HI} provided
\begin{gather*}
 \mytH{\xi}^{2}\mytI{\xi} = \frac{ 1 }{ \xi + \xi^{-1} - \nu_{1} - \nu_{2} }.
\end{gather*}

Thus, to obtain solutions for the equations we are to solve, we have just to return to the $\n$-dependence knowing the action of~$\myShift{\xi}$. To this end, we have to take $\xi \in \{ g_{1}, g_{2}, g_{3} \}$ where~$g_{i}$ (parameters that, recall, correspond to the vectors $\myg{i}$) should be determined from
\begin{gather}\label{dark:g}
 g_{i} + g_{i}^{-1} = \nu_{1} + \nu_{2} + \frac{ 1 }{ H_{i}^{2}\myI{i} }
\end{gather}
with constants $\myH{i}$ and $\myI{i}$ ($i=1,2,3$) (we write $\myH{i}$ and $\myI{i}$ instead of~$\mytH{g_{i}}$ and $\mytI{g_{i}}$) defined in~\eqref{rslt:H} and~\eqref{rslt:I}.

\begin{Proposition}The `dark-soliton' solutions for the field equations can be presented as
\begin{gather*}
 \vq\myarg{} =
 \frac{ 1 }
 {\det | \mymatrix{1} + \mymatrix{A}\myarg{} | }
 \left(\begin{matrix}
 q_{1} e^{ \mysp{ \vec{\varphi}_{1} }{ \n } }\det | \mymatrix{1} + \mymatrix{A}\myarg{} \mymatrix{T}_{\nu_{1}}^{-1}| \\
 q_{2} e^{ \mysp{ \vec{\varphi}_{2} }{ \n } }
\det| \mymatrix{1} + \mymatrix{A}\myarg{} \mymatrix{T}_{\nu_{2}}^{-1}| \end{matrix}\right)
\end{gather*}
and
\begin{gather*}
 \vr\myarg{} = \frac{ 1 }
 {\det | \mymatrix{1} + \mymatrix{A}\myarg{} | }
 \left(\begin{matrix}
 r_{1} e^{ - \mysp{ \vec{\varphi}_{1} }{ \n } }
\det | \mymatrix{1} + \mymatrix{A}\myarg{} \mymatrix{T}_{\nu_{1}}| \\
 r_{2} e^{ - \mysp{ \vec{\varphi}_{2} }{ \n } }
\det| \mymatrix{1} + \mymatrix{A}\myarg{} \mymatrix{T}_{\nu_{2}}| \end{matrix}\right),
\end{gather*}
where $q_{a}$ and $r_{a}$ are defined in~\eqref{dark:cqr}, the vectors $\vec{\varphi}_{a}$ are given by
\begin{gather*}
 \vec{\varphi}_{a} = \frac{2}{3} \sum_{i=1}^{3} \left( \ln \frac{ \left( g_{i-1} - \nu_{a} \right) \myH{i-1} } { \left( g_{i+1} - \nu_{a} \right) \myH{i+1} } \right) \mye{i},
 \qquad a=1,2,
\end{gather*}
while the matrix $\mymatrix{A}\myarg{}$ is given by
\begin{gather*}
 \mymatrix{A}\myarg{} = \mymatrix{A}_{0} \operatorname{diag} \big( \exp \mysp{ \vec{\phi}_{\ell} }{ \n } \big)_{\ell=1}^{N},
\end{gather*}
where
\begin{gather*}
 \vec{\phi}_{\ell} = \frac{2}{3} \sum_{i=1}^{3} \left( \ln \frac{ g_{i+1} - L_{\ell} }{ g_{i+1} - L_{\ell}^{-1} } \frac{ g_{i-1} - L_{\ell}^{-1} }{ g_{i-1} - L_{\ell} } \right) \mye{i}
\end{gather*}
and
\begin{gather*}
 \mymatrix{A}_{0} = \left( \frac{ a_{k} }{ 1 - L_{j}L_{k} } \right)_{j,k=1}^{N}.
\end{gather*}
Here, the parameters $g_{1}$, $g_{2}$ and $g_{3}$ should be determined from~\eqref{dark:g}, while $L_{k}$, $a_{k}$ $(k=1, \dots , N)$, $\nu_{1}$, $q_{1}$, $q_{2}$ as well as $\myH{*}$ and $\myI{*}$ in~\eqref{rslt:H} and~\eqref{rslt:I} are arbitrary constants.
\end{Proposition}

\subsection{`Bright' solitons}

To derive the second type of soliton solutions one can use the soliton Fay identities from \cite{V15} which were obtained for the tau-functions
\begin{gather}\label{bright:tau}
 \tau = \det | \mymatrix{1} + \mymatrix{A}\mymatrix{B} |,
\qquad
 \sigma = \tau \big\langle a | ( \mymatrix{1} + \mymatrix{B}\mymatrix{A} )^{-1} | \beta \big\rangle,
\qquad
\rho = \tau \big\langle b | ( \mymatrix{1} + \mymatrix{A}\mymatrix{B} )^{-1} | \alpha \big\rangle,
\end{gather}
where $(N \times N)$-matrices $\mymatrix{A}$ and $\mymatrix{B}$ are solutions of
\begin{gather*}
 \mymatrix{L} \mymatrix{A} - \mymatrix{A} \mymatrix{R} = | \alpha \rangle \langle a |,
\qquad
 \mymatrix{R} \mymatrix{B} - \mymatrix{B} \mymatrix{L} = | \beta \rangle \langle b |.
\end{gather*}
Here, like in the previous section, $\mymatrix{L}$ and $\mymatrix{R}$ are constant diagonal $(N \times N)$-matrices, $\mymatrix{L} = \operatorname{diag} ( L_{1}$, $\dots,L_{N} )$ and $\mymatrix{R} = \operatorname{diag} ( R_{1}, \dots , R_{N} )$, $| \alpha \rangle$ and $| \beta \rangle$ are constant $N$-columns, $\langle a |$ and $\langle b |$ are $N$-component rows that depend on the coordinates describing the model.

The shifts $\myShift{\xi}$ are defined, in this case, by
\begin{gather*}
 \myShifted{\xi}{\langle a |} = \langle a | (\mymatrix{R} - \xi)^{-1},\qquad
 \myShifted{\xi}{\langle b |} = \langle b | (\mymatrix{L} - \xi)
\end{gather*}
or, as a consequence, by
\begin{gather*}
 \myShifted{\xi}{\mymatrix{A}} = \mymatrix{A} (\mymatrix{R} - \xi)^{-1},
\qquad \myShifted{\xi}{\mymatrix{B}} = \mymatrix{B} (\mymatrix{L} - \xi).
\end{gather*}

The simplest soliton identities from \cite{V15} are
\begin{gather}\label{bright:bist}
(\xi - \eta) \tau \myShifted{\xi\eta}{\sigma} = \myShifted[1]{\xi}{\sigma} \myShifted[1]{\eta}{\tau} - \myShifted[1]{\xi}{\tau} \myShifted[1]{\eta}{\sigma},\\
(\xi - \eta) \rho \myShifted{\xi\eta}{\tau} = \myShifted[1]{\xi}{\tau} \myShifted[1]{\eta}{\rho} - \myShifted[1]{\xi}{\rho} \myShifted[1]{\eta}{\tau}
\end{gather}
and
\begin{gather}
 \myShifted[1]{\xi}{\tau} \myShifted[1]{\eta}{\tau} = \tau \myShifted{\xi\eta}{\tau} + \rho \myShifted{\xi\eta}{\sigma}.\label{bright:alh}
\end{gather}

The fact that $\tau$, $\sigma$ and $\rho$ are solutions of \eqref{bright:bist}--\eqref{bright:alh} is enough to demonstrate that the vectors $\vec{\sigma}$ and $\vec{\rho}$ given by
\begin{gather*}
 \vec{\sigma} = \left( \begin{matrix}
\myE_{1} \myShiftInv{\kappa} \tau
\\[1mm]
 \myE_{2} \myShift{\kappa} \sigma
\end{matrix}
\right),
\qquad \vec{\rho} = \left( \begin{matrix}\myE_{1}^{-1}\myShift{\kappa} \tau\\[1mm]
\myE_{2}^{-1}\myShiftInv{\kappa} \rho\end{matrix}\right),
\end{gather*}
where $\myE_{1}$ and $\myE_{2}$ are defined by
\begin{gather*}
\myShifted{\xi}{\myE_{1}} = \mytH{\xi}\mytI{\xi} \myE_{1}, \qquad
\myShifted{\xi}{\myE_{2}} = (\kappa - \xi) \mytH{\xi}\mytI{\xi} \myE_{2}
\end{gather*}
with arbitrary $\kappa$ satisfy equations \eqref{bi:s} and \eqref{bi:r} with
\begin{gather*}
 A_{\xi,\eta} = (\eta - \xi) \mytH{\xi}\mytH{\eta}
\end{gather*}
as well as equations \eqref{bi:u} and \eqref{bi:HI} provided
\begin{gather*}
 \xi - \kappa = \frac{ 1 }{ \mytH{\xi}^{2}\mytI{\xi} }.
\end{gather*}

To simplify the final formulae, we introduce the diagonal matrices $\mymatrix{X}\myarg{}$ and $\mymatrix{Y}\myarg{}$, describing the $\n$-dependence of the rows $\langle a\myarg{} |$ and $\langle b\myarg{} |$,
\begin{gather*}
 \langle a\myarg{} | = \langle a | \mymatrix{X}\myarg{}, \qquad
 \langle b\myarg{} | = \langle b | \mymatrix{Y}\myarg{},
\end{gather*}
as well as the new rows $\langle \hat{a}\myarg{} |$ and $\langle \hat{b}\myarg{} |$ defined as
\begin{gather}
 \langle \hat{a}\myarg{} | = \langle a | \big[ \mymatrix{X}^{-1}\myarg{} + \mymatrix{B} \mymatrix{Y}\myarg{} \mymatrix{A} \big]^{-1},\nonumber \\
 \langle \hat{b}\myarg{} | = \langle b | \big[ \mymatrix{Y}^{-1}\myarg{} + \mymatrix{A} \mymatrix{X}\myarg{} \mymatrix{B} \big]^{-1}.\label{bright:hbra}
\end{gather}

After some simple transformations of the matrix formulae \eqref{bright:tau} and elimination of some `redundant' constants (which includes setting $\kappa = 0$) we arrive at the following result.

\begin{Proposition}The `bright-soliton' solutions for the field equations can be presented as
\begin{gather*}
 \vq\myarg{} = \left(
 \begin{matrix}
 C_{1}e^{ \mysp{ \vec{\varphi}_{1} }{ \n } }\left[ 1 - \langle \hat{a}\myarg{} | \mymatrix{B} \mymatrix{Y}\myarg{} | 1 \rangle \right] \\[1mm]
C_{2}e^{ \mysp{ \vec{\varphi}_{2} }{ \n } }\langle \hat{a}\myarg{}| 1 \rangle\end{matrix}\right),\\
 \vr\myarg{} = \left( \begin{matrix}
 C_{1}^{-1} e^{ - \mysp{ \vec{\varphi}_{1} }{ \n } }\big[ 1 - \langle \hat{b}\myarg{} | \mymatrix{A}
 \mymatrix{X}\myarg{} | 1 \rangle \big]\\[1mm]
 C_{2}^{-1} e^{ - \mysp{ \vec{\varphi}_{2} }{ \n } }\langle \hat{b}\myarg{} | 1 \rangle\end{matrix}\right),
\end{gather*}
where
\begin{gather*}
 \vec{\varphi}_{1} = \frac{2}{3} \sum_{i=1}^{3} \left( \ln \frac{ \myH{i+1}\myI{i+1} }{ \myH{i-1}\myI{i-1} } \right) \mye{i},\qquad
 \vec{\varphi}_{2} = \frac{2}{3} \sum_{i=1}^{3} \left( \ln \frac{ \myH{i-1} }{ \myH{i+1} } \right) \mye{i},
\end{gather*}
the rows $\langle \hat{a}\myarg{} |$ and $\langle \hat{b}\myarg{} |$ are defined in~\eqref{bright:hbra} with
\begin{gather*}
 \mymatrix{X}\myarg{} = \operatorname{diag} \big( \exp \mysp{ \vec{\phi}_{\ell} }{ \n } \big)_{\ell=1}^{N},
\qquad \mymatrix{Y}\myarg{} = \operatorname{diag} \big( \exp \mysp{ \vec{\psi}_{\ell} }{ \n } \big)_{\ell=1}^{N}
\end{gather*}
and
\begin{gather*}
 \vec{\phi}_{\ell} = \frac{2}{3} \sum_{i=1}^{3} \left( \ln \frac{ R_{\ell} - g_{i-1}}{ R_{\ell} - g_{i+1} } \right) \mye{i},
\qquad \vec{\psi}_{\ell} = \frac{2}{3} \sum_{i=1}^{3} \left( \ln \frac{ L_{\ell} - g_{i+1} }{ L_{\ell} - g_{i-1} } \right) \mye{i}, \qquad
 g_{i} = \frac{ 1 }{ \myH{i}^{2}\myI{i} },
\end{gather*}
$| 1 \rangle$ is the $N$-column with all elements equal to $1$, $\langle a |$ and $\langle b |$ are constant $N$-rows, $\langle a | = ( a_{1}, \dots , a_{N} )$, $\langle b | = ( b_{1}, \dots , b_{N})$ and constant matrices $\mymatrix{A}$ and $\mymatrix{B}$ are given by
\begin{gather*}
 \mymatrix{A} = \left( \frac{ L_{j} a_{k} }{ L_{j}-R_{k} } \right)_{j,k=1}^{N},\qquad \mymatrix{B} = \left( \frac{ R_{j} b_{k} }{ R_{j}-L_{k} } \right)_{j,k=1}^{N}.
\end{gather*}
Here, $C_{1}$, $C_{2}$, $L_{\ell}$, $R_{\ell}$, $a_{\ell}$, $b_{\ell}$ $(\ell=1, \dots , N)$ as well as $\myH{*}$ and $\myI{*}$ in \eqref{rslt:H} and \eqref{rslt:I} are arbitrary constant parameters of the solution.
\end{Proposition}

\section{Conclusion}

In this paper we have considered the vector model on the triangular lattice. It should be noted that we have not elaborated some special methods to take into account the vector character of the model or the fact that the lattice is not a rectangular one. Instead, we have used simple algebraic calculations to demonstrate that this somewhat non-standard model can be reduced to the already known equations which are usually associated with the rectangular lattices. Of course, this is a reduction. But this reduction is not a trivial one, in the sense that it leads to a~rather wide range of solutions, which includes not only solutions presented above but also the so-called finite-gap quasiperiodic and many other solutions (not discussed here). Thus, one of the ideas behind this work is that there are much more soliton models than in the `standard' set of the integrable ones.

However, in doing this we have met the manifestations of the peculiarities of the triangular geometry. First of all, one should mention the rather intricate relations between the constants and the special role of the anisotropy: note that~\eqref{restr:JK} implies that one cannot take $\myJ{1}=\myJ{2}=\myJ{3}$ or $\myK{1}=\myK{2}=\myK{3}$, which can be viewed as some kind of frustration already known to occur in the triangular lattices.
Although the question of whether the restrictions~\eqref{restr:JK}, which are crucial for our calculations, are necessary for existence of solitons is an open one, one cannot deny the importance of the anisotropy in this
model. Comparing the soliton solutions of this paper with ones derived in~\cite{V16b} for the 3D vectors (instead of 2D) with interaction similar to~\eqref{mod:L}, but isotropic, and without the restriction \eqref{restr:qr} one can conclude that it is interesting to study the interplay between the dimensionality of the vectors, anisotropy and the length restrictions.

Finally, we would like to mention the following limitations of the ansatz used in this paper. The case is that the model \eqref{mod:S} with \eqref{mod:L} possesses some reductions that are rather interes\-ting from the viewpoint of applications. The simplest one is $\vq = \vr$, which after resolving the restriction~\eqref{restr:qr}, $|\vq\,|^{2} = 1$, by presenting $\vq$ as $\vq = ( \cos\theta, \sin\theta )^{\scriptscriptstyle\rm T}$
leads to
\begin{gather*}
 \sum\limits_{i=1}^{3} \frac{ \myJ{i}\bar{K}_{i} \sin [ \theta\myarg{} - \theta\myarg{+\mye{i}} ] }{ 1 - \bar{K}_{i} \cos [ \theta\myarg{} - \theta\myarg{+\mye{i}} ] } = 0,
\end{gather*}
where $\bar{K}_{i} = \myK{i}/ (\myK{i} + 1/2 )$. Unfortunately, all solutions presented in this paper become trivial under this reduction. The problem is that introducing the auxiliary system \eqref{eq:aux-syst} we break the symmetry between $\vq$ and $\vr$ inherent in the system \eqref{eq:main-q} and \eqref{eq:main-r}. In other words, our ansatz is incompatible with the reduction $\vr \to \vq$. In a similar way, our ansatz is also incompatible with another reduction of physical interest, $\vr \to \vq^{\,*}$ where the star indicates the complex conjugation. Thus, a natural continuation of this work is to replace the `triangular' system \eqref{eq:aux-syst} with another one, say, a system of quad-equations having the symmetries discussed above.

However, these questions are out of the scope of the present paper and may be addressed in future studies.

\appendix

\section{Consistency of (\ref{eq:aux-syst}) and proof of (\ref{eq:U})}\label{app:a}

Here, we derive the restrictions \eqref{eq:J} and \eqref{eq:K} and prove the fact that the quantity
\begin{gather*}
 U_{i}\myarg{} = \mysp{ \vq\myarg{+\myg{i}} }{ \vr\myarg{} }
\end{gather*}
is constant or, that $U_{i}\myarg{+\myg{j}} = U_{i}\myarg{}$ for all $i$ and $j$ ($i,j=1,2,3$).

Subtracting the first equation of \eqref{eq:aux-syst} multiplied by $\vr\myarg{+g_{i-1}}$ from the second equation of~\eqref{eq:ansatz} multiplied by $\vq\myarg{+g_{i+1}}$ one obtains
\begin{gather*}
 U_{i+1}\myarg{+\myg{i-1}} = U_{i+1}\myarg{}.
\end{gather*}
Repeating this procedure for different values of the indices, one can obtain
\begin{gather}
 U_{i}\myarg{+\myg{j}} = U_{i}\myarg{}, \qquad i \ne j.\label{appa:a}
\end{gather}
Next, after multiplying the second equation of~\eqref{eq:aux-syst} by $\vr\myarg{+g_{i-1}}$ and $\vr\myarg{+g_{i+1}}$ one can obtain
\begin{gather}
 \myH{i-1} \mysp{ \vq\myarg{+\myg{i-1}} }{ \vr\myarg{+\myg{i+1}} } - \myH{i+1} = \frac{ 1 }{ \myJ{i} } F_{i}\myarg{} U_{i-1}\myarg{}\label{appa:b}
\end{gather}
and
\begin{gather}
 \myH{i-1} - \myH{i+1} \mysp{ \vq\myarg{+\myg{i+1}} }{ \vr\myarg{+\myg{i-1}} } = \frac{ 1 }{ \myJ{i} } F_{i}\myarg{} U_{i+1}\myarg{}.
\label{appa:c}
\end{gather}
Substitution of the scalar products from \eqref{appa:b} and \eqref{appa:c} into the definition of $F_{i}\myarg{}$, which can be written as
\begin{gather*}
 F_{i}\myarg{} = \frac{ 1 }{ \myK{i} } + 2 - \mysp{ \vq\myarg{+\myg{i-1}} }{ \vr\myarg{+\myg{i+1}} } - \mysp{ \vq\myarg{+\myg{i+1}} }{ \vr\myarg{+\myg{i-1}} },
\end{gather*}
leads to
\begin{gather*}
 \mathcal{A}_{i}\myarg{} F_{i}\myarg{} = \mathcal{B}_{i},
\end{gather*}
where
\begin{gather}
 \mathcal{A}_{i}\myarg{} = 1 + \frac{ 1 }{ \myJ{i} } \frac{ U_{i-1}\myarg{} }{ \myH{i-1} } - \frac{ 1 }{ \myJ{i} } \frac{ U_{i+1}\myarg{} }{ \myH{i+1} }\label{appa:d}
\end{gather}
and
\begin{gather*}
 \mathcal{B}_{i} = \frac{ 1 }{ \myK{i} } + 2 - \frac{ \myH{i-1} }{ \myH{i+1} } - \frac{ \myH{i+1} }{ \myH{i-1} }.
\end{gather*}
Noting that $\mathcal{A}_{i}\myarg{+\myg{i}} = \mathcal{A}_{i}\myarg{}$, because of \eqref{appa:d} and \eqref{appa:a}, one can conclude that either $F_{i}\myarg{+\myg{i}} = F_{i}\myarg{}$ (we do not consider this possibility as leading to rather trivial solutions) or
\begin{gather*}
 \mathcal{A}_{i}\myarg{} = \mathcal{B}_{i} = 0.
\end{gather*}
The second of these equalities,
\begin{gather*}
 \frac{ 1 }{ \myK{i} } = \frac{ \myH{i-1} }{ \myH{i+1} } + \frac{ \myH{i+1} }{ \myH{i-1} } - 2,
\end{gather*}
is nothing but \eqref{eq:K}. The first one, $\mathcal{A}_{i} = 0$, leads, after the translations $\n \to \n + \myg{i \pm 1}$, to $U_{i}\myarg{+\myg{i}} = U_{i}\myarg{}$ which, together with~\eqref{appa:a}, means that
\begin{gather*}
 U_{i}\myarg{+\myg{j}} = U_{i}\myarg{}, \qquad i,j=1,2,3,
\end{gather*}
i.e., that $U_{i}\myarg{}$ are constants with respect to $\n$,
\begin{gather*}
 U_{i}\myarg{} = U_{i} = {\rm const}, \qquad i=1,2,3.
\end{gather*}
Introducing the new constants, $\myI{i} = U_{i} / \myH{i}$, one can obtain from~\eqref{appa:d} the constraint~\eqref{eq:J}.

\subsection*{Acknowledgments}
We would like to thank the referees for their constructive comments and suggestions for improvement of this paper.

\pdfbookmark[1]{References}{ref}
\LastPageEnding

\end{document}